\documentclass[prd,twocolumn,superscriptaddress,showpacs,amsmath,amssymb]{revtex4}
\usepackage{graphicx}
\usepackage{dcolumn}
\usepackage{bm}
\usepackage{epsfig}
\usepackage{amsfonts}
\begin{document}

\title{Self-organized critical behavior: the evolution of frozen spin networks model in quantum gravity}

\author{Jian-Zhen Chen}
  \email{jzhchan@yahoo.com}
  \affiliation{Department of Physics, Beijing Normal University, Beijing 100875, China}
  \affiliation{Department of Physics, Jiangxi Normal University, Nanchang 330027, China}

\author{Jian-Yang Zhu}
\thanks{Author to whom correspondence should be addressed}
  \email{zhujy@bnu.edu.cn}
  \affiliation{Department of Physics, Beijing Normal University, Beijing 100875, China}
\date{\today}

\begin{abstract}
In quantum gravity, we study the evolution of a two-dimensional
planar open frozen spin network, in which the color (i.e. the twice
spin of an edge) labeling edge changes but the underlying graph
remains fixed. The mainly considered evolution rule, the random edge
model, is depending on choosing an edge randomly and changing the
color of it by an even integer. Since the change of color generally
violate the gauge invariance conditions imposed on the system,
detailed propagation rule is needed and it can be defined in many
ways. Here, we provided one new propagation rule, in which the
involved even integer is not a constant one as in previous works,
but changeable with certain probability. In random edge model, we do
find the evolution of the system under the propagation rule exhibits
power-law behavior, which is suggestive of the self-organized
criticality (SOC), and it is the first time to verify the SOC
behavior in such evolution  model for the frozen spin network.
Furthermore, the increase of the average color of the spin network
in time can show the nature of inflation for the universe.
\end{abstract}

 \maketitle

\section{Introduction}
\label{Sec.1}

Self-organized critical phenomenon, is exhibited by a dynamic system
that displays spatial/temporal scale invariance characteristic of
the critical point of phase transition, but without the need of
external fine tuning of control parameters \cite{Bak}. Many such
examples can be found in natural and social systems, including
granular materials, earthquakes, biological evolution and stock
markets \cite{Bak,Bak1}.

Recently, the self-organized criticality (SOC) also became an
attractive problem in quantum gravity
\cite{Markopoulou,Borissov,Ansari}. Rovelli and Smolin
\cite{Rovelli1} pointed out that the space has discrete structure at
Plank scale in canonical quantum gravity
\cite{Ashtekar,Rovelli,Smolin}, and then it was proposed by
Markopoulou et al. \cite{Markopoulou} that the discrete space should
be a SOC system for the lack of an external fine tuning to the
universe's parameters. The proposal was studied further by Borissov
and Gupta \cite{Borissov}, and then by Ansari and Smolin
\cite{Ansari} more recently. For other researches on critical
phenomenon in quantum gravity see refs. \cite{Critical,Critical2}.

The research on the SOC in quantum gravity \cite{Borissov,Ansari}
was based on the frozen spin network, a simplified case of loop
quantum gravity, in which the embedding graph is fixed and the spin
that labels the edge evolves. Furthermore, the identities of the
gauge invariance  imposed at vertices is not chosen as the
conditions on states, but a dynamical process in which the system
evolves to gauge invariance state. The evolution rules adopted by
Borissov and Gupta \cite{Borissov} did not lead to the SOC. Then
Ansari and Smolin \cite{Ansari} developed a new set of evolution
rules and observed the SOC. The authors mainly applied the classical
statistical mechanics on spin networks instead of quantum mechanics,
and the quantum SOC is still a significant issue to be studied.

In this paper, we developed a set of more realistic propagation
rules based on the ideas of refs. \cite{Borissov,Ansari}, and
studied the SOC in loop quantum gravity. In Sec. \ref{Sec.2}, the
model of spin network adopted in this paper is briefly reviewed.
Following that, the two main classes of evolution rules for the
frozen spin networks are described in Sec. \ref{Sec.3}. One is the
random edge model, which involves changing the color (i.e. the twice
spin of an edge) of a randomly chosen edge by an even integer. The
other one is the random vertex model, which involves choosing a
vertex at random and evolving the colors of its three incident edges
by an even integer. Since the changes of color generally lead to the
violation of the gauge invariance conditions, detailed propagation
rules should be defined for the spin network to regain the gauge
invariance. The propagation rules can be defined in different ways.
In Sec. \ref{Sec.4} we presented a new possible propagation rule, in
which the change of the edge color is an even integer chosen with a
certain probability distribution, but no longer a fixed value as in
previous works. We found that the system under our propagation rule
exhibits the SOC in the random edge model for the first time.
Summary and discussions can be found in Sec. \ref{Sec.5}.

\section{The spin network}

\label{Sec.2}

As introduced by Rovelli and Smolin \cite{Rovelli2}, and then by
Baez \cite{Baez}, in canonical quantum gravity, a spin network state
is defined by a closed graph $\gamma$ which is constructed by a
finite number of oriented edges $e_{1},e_{2}\cdots$ incident at
vertices $v_{1},v_{2}\cdots$. A vertex with $p$ incident edges is
called a $p$-valent vertex or a vertex of valence $p$. The edge
$e_{i}$ is labeled by a irreducible representation $j_{i}$ (i.e.
spin) of SU(2). The color of edge $e_{i}$ is defined as its twice
spin, $c_{i}=2j_{i}$.

\begin{figure}
\includegraphics[width=0.35\textwidth]{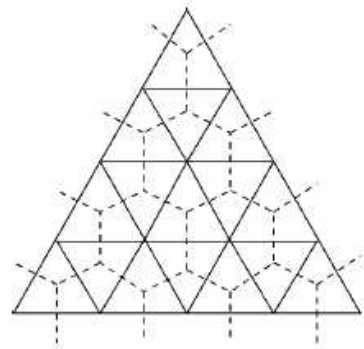}
\caption{The triangulated triangle (solid lines) and its dual,
 trivalent spin network (dashed lines).}
\label{Fig1}
\end{figure}

For simplicity, we consider only the planar trivalent spin network
which is dual to a colored triangulation of the space
\cite{Markopoulou1}. Firstly, we take a triangular space and divide
it into $N$ sub-triangles to perform the triangulation. The result
of the triangulation is exhibited by the solid lines in Fig.
\ref{Fig1}. Then the dual trivalent spin network is constructed by
connecting the centers of each sub-triangle to the centers of its
adjacent sub-triangles, and its boundary is given by the dual to the
segments of the edges of the original triangle space. The dual spin
network is shown in Fig. \ref{Fig1} as the dashed lines.

The length of a side in the triangulated network and the color of
its dual edge in the spin network have the following relationship,
$2l_{side}=l_{Plank}\cdot c_{edge}$ \cite{Rovelli3}. Since the
lengths of the three sides of a triangle obey the triangle
inequalities, the colors of edges incident at a vertex in the dual
spin network is constrained by the following \emph{gauge invariance
condition}: (1) the triangle inequalities are satisfied and (2) the
sum of the colors of a vertex's incident edges should be an even
integer. Suppose $a,b$ and $c$ are the colors of edges incident at a
vertex and they are all positive integers, the gauge invariance
condition for a vertex can be expressed as
\begin{equation}
a+b\geqslant c, b+c\geqslant a, c+a\geqslant b; \label{gnc1}
\end{equation}
\begin{equation}
a+b+c=even. \label{gnc2}
\end{equation}
If the gauge invariance condition (eqs. (\ref{gnc1}, \ref{gnc2})) is
not satisfied for a vertex, that vertex is called a
gauge-non-invariance (GNI) vertex. The spin network reaches a gauge
invariance state when the gauge invariance condition is satisfied at
all vertices.

The spin network evolves from one gauge invariance state to another
one by adding or subtracting a loop flux. The evolving rules fall
into two main classes \cite{Ansari,Borissov}: One is the random
vertex class, which involves choosing a random vertex and
simultaneously changing the colors of its three incident edges by an
even integer $\triangle c$. The other one is the random edge class
in which one edge is chosen randomly and its color is changed by an
even number $\triangle c$. The two evolution rules will be
detailedly discussed in Sec. \ref{Sec.3} respectively.

The spin network studied hereafter is constructed using the method
described above (Fig. \ref{Fig1}). It is a trivalent spin network in
a two-dimensional space with open boundary. Moreover, during the
evolution of the spin network, the underlying graph is fixed and
only the color of the edges can be changed. Such model is also
referred to as \emph{the frozen spin network}. In the initial state,
random integers between $1$ and $c_{max}$, where $c_{max}$ is a
given maximum value of the initial color, are assigned to each edge
and the spin network is required to be gauge invariant. For an
easier comparison with the previous works, $c_{max}$ is chosen to be
$30$ as in Ref. \cite{Ansari} in the following parts unless noted
specifically. But we can see in Sec. \ref{Sec.5} that the values of
$c_{max}$ have negligible influence on the SOC behavior.

\section{The random edge model and the random vertex model}

\label{Sec.3}

The random edge model: In \cite{Borissov}, Borissov and Gupta
established their evolution rule on a two-dimensional planar spin
network. (1) Initialize the graph. (2) Choose an edge at random and
change its color by $\triangle c =-2$. This will generally break the
gauge invariance condition of the two vertices linked by that edge.
(3) To restore the gauge invariance state, the system evolves with a
propagation rule, which can be defined in many different ways (see
\cite{Ansari, Borissov} for examples). When the gauge invariance is
restored, a new edge is selected and step (3) is repeated.

The random vertex model: After initializing the spin network, the
propagation rule of the random vertex model, according to Ansari and
Smolin in \cite{Ansari}, consists of the following steps: (1) Choose
a vertex randomly and subtract 2 from the color of each of its
edges; (2) Check the gauge condition (\ref{gnc1},\ref{gnc2}) at all
vertices. If the gauge invariance does not hold for a vertex, then
add $\triangle c^{\prime}=+2$ to the color of each of its edges.
Continue untill the gauge invariance is restored, and then repeat
steps (1) and (2).

The random edge model was studied in Ref. \cite{Borissov}, and the
SOC was not observed. Ref. \cite{Ansari} studied the random vertex
model and found to display the SOC behavior.

In the following, we will perform the work of our detailed
propagation rule in the two main models respectively. Our
propagation rule can verify the SOC behavior in random edge model,
but not in random vertex model. Therefore we will focus on the
discussion of the random edge model in the next section.

\section{Our propagation rule and the results}

\label{Sec.4}

Reviewing the random edge model, one may find that the main steps in
the evolution of the spin network can be summarized as the
following: First, construct and initialize the graph. Then, (1)
randomly choose one edge and subtract the color of it by $\triangle
c$ \footnote{In our paper, the randomly chosen edge should not have
color less than $\triangle c$, otherwise it will be skipped with no
action taken.}. (2) If the gauge invariance does not hold for a
vertex, randomly choose one of its incident edges and add $\triangle
c^{\prime}$ to its color. This is repeated until the graph is in
gauge invariance state again. The set of consecutive updates of
vertices in recovering the gauge invariance  is an avalanche, and
the number of updated vertices is the size of the avalanche. The
area of the avalanche has a different definition: it is the number
of vertices involved in one avalanche, no matter how many times they
are updated. (3) Repeat steps (1) and (2) for a large number of
times to see the behavior of the spin network in a long time and
hunt for the propagation rules to lead the distribution of size of
avalanches become scale variant.

In \cite{Ansari}, the authors initialized the network by assigning
to each edge a random even integer between $10$ and $30$. Here we
extend the initialization to a more general condition. The initial
colors are randomly selected from the integers between $1$ and
$c_{max}$. Furthermore, the $\triangle c$ involved in step (1) is
not fixed as $2$ any more, but an even integer randomly chosen
between $2$ and $c_{max}$.

The more important generalization is in step (2). Since the vertex
is not isolated, but affected by the adjacent vertices in general,
the $\triangle c^{\prime}$ involved in the update of a GNI vertex is
suggested to be a variable dependent on the original color of the
vertex and the colors of its adjacent vertices jointly. Assume the
GNI vertex in study, $v_{i}$, has three adjacent vertices $v_{1}$,
$v_{2}$ and $v_{3}$ \footnote{If the number of the actual adjacent
vertices is less than three, the edges' color of the nonexistent
adjacent vertices are set to be zero.}, the sums of color of the
vertices' three incident edges are $c_{i}$, $c_{1}$, $c_{2}$ and
$c_{3}$ respectively. During the update, the color of each of the
incident edges on vertex $v_{i}$ is changed by $\triangle
c^{\prime}_{i}$. $c^{\prime}_{i}$ is an even integer chosen between
$2$ and a maximum value, $\triangle c_{i,max}^{\prime}$, with a
probability distribution,

\begin{equation}
p=\frac{\exp(-\beta \triangle c_{i}^{\prime})}{\sum_{\triangle
c_{i}^{\prime}}\exp(-\beta \triangle c_{i}^{\prime})}, \notag
\end{equation}
where
\begin{equation}
\beta =\frac{1}{|(c_{1}+c_{2}+c_{3})/3-c_{i})|} \notag
\end{equation}  and
\begin{equation}
\triangle c_{i}^{\prime}=2,4,\cdots,\triangle c_{i,max}^{\prime}.
\notag
\end{equation}

When $(c_{1}+c_{2}+c_{3})/3=c_{i}$, we take $p=1$ when $\triangle
c_{i}^{\prime}=2$ and $p=0$ for other values of $\triangle
c_{i}^{\prime}$.

In the present study, $\triangle c^{\prime}_{i,max}$ is set to be
equal to the maximum value of the initialized color, $c_{max}$ (or
$c_{max}-1$ if $c_{max}$ is an odd number). Obviously, $\triangle
c^{\prime}_{i,max}$ can have other suggestions, but for the sake of
brevity, the other suggestions will not be discussed here.

\begin{figure}
\includegraphics[clip,width=0.38\textwidth]{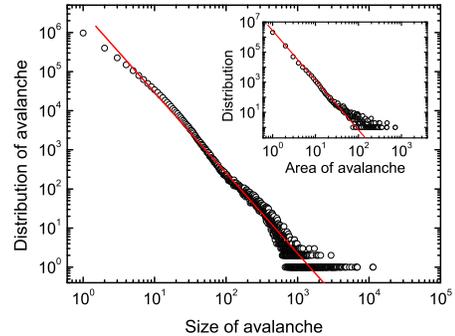}
\caption{(Color online) Results for a typical run in ten million
iterations in the random edge model for an open planar spin network
with 102400 vertices. The graph shows the log-log distribution of
avalanche, and the fitted line shows the power-law relation:
$P(s)\propto s^{-2.06}$. The corresponding distribution of area is
given in the log-log subplot.} \label{Fig.2}
\end{figure}

\begin{figure}
\includegraphics[clip,width=0.42\textwidth]{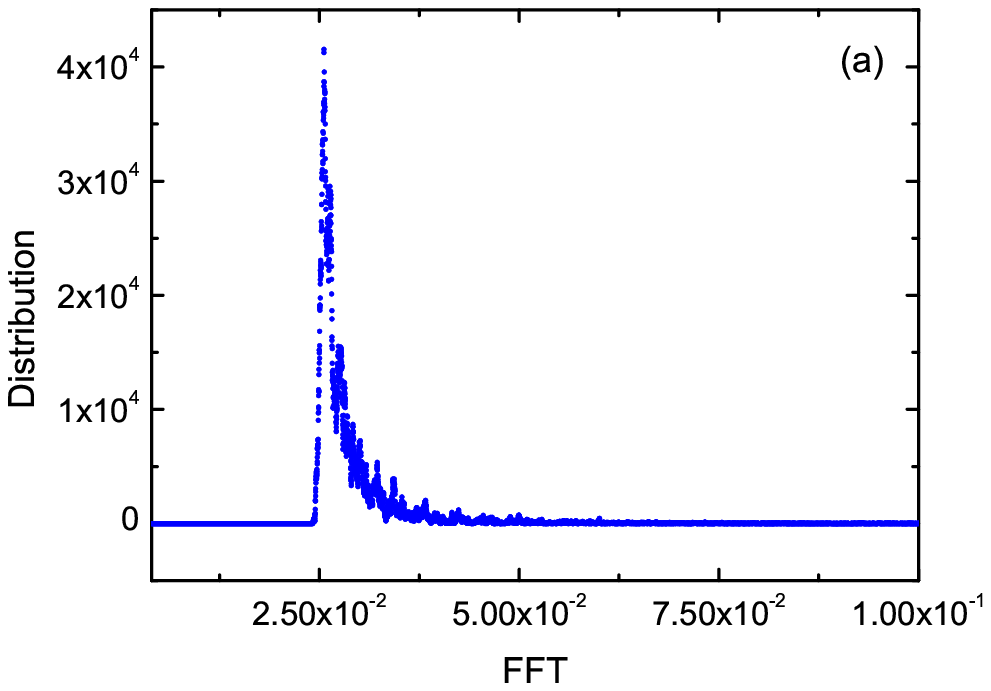}
\includegraphics[clip,width=0.45\textwidth]{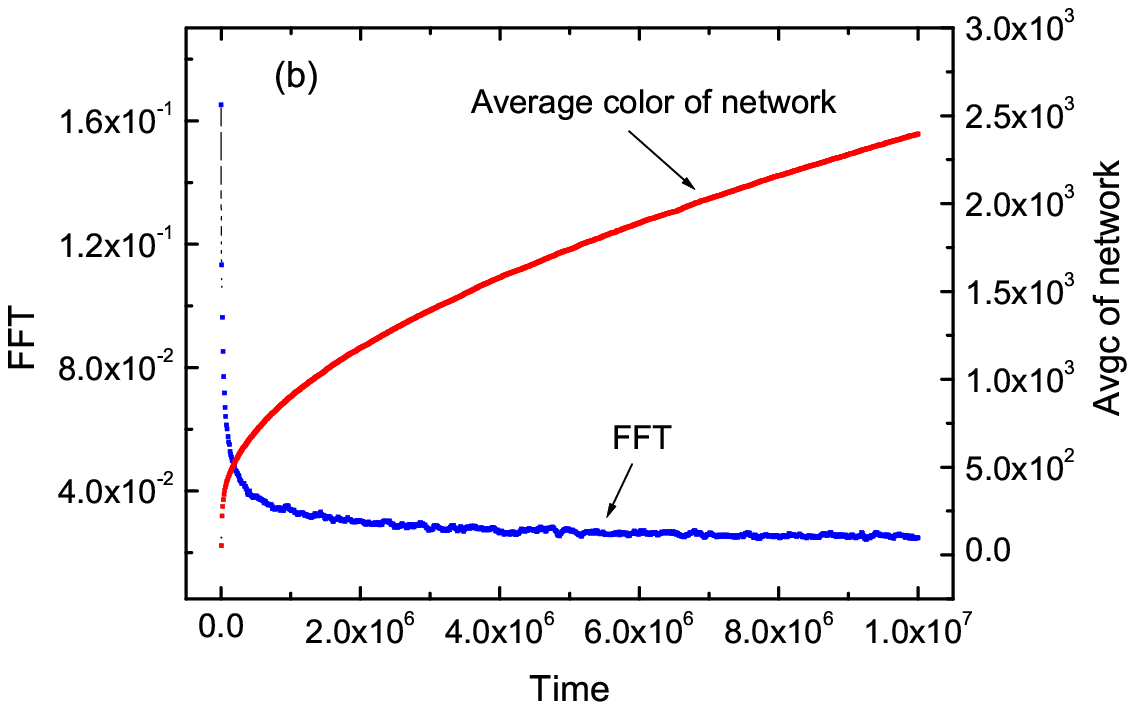}
\caption{(Color online) Results for a typical run in ten million
iterations in the random edge model for an open planar spin network
with 102400 vertices. The distribution of the FFT is provided in
Graph (a). Graph (b) shows the fraction of the flat triangles (abbr.
FFT) and the average color (abbr. Avgc) of spin network in time
respectively. }\label{Fig.3}
\end{figure}

We did the simulation for $10^7$ steps on a spin network with 102400
vertices. The results of a typical run are shown in Figs.
\ref{Fig.2} and \ref{Fig.3}.

\begin{figure}[tbp]
\includegraphics[clip,width=0.38\textwidth]{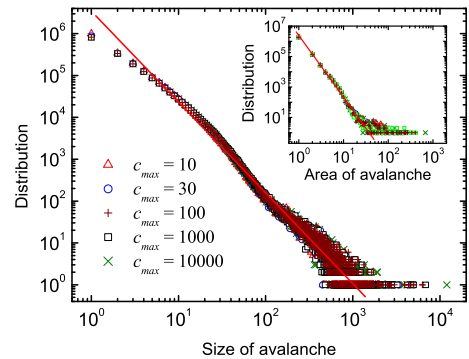}
\caption{(Color online) Results of typical runs in ten million
iterations in the random edge model on an open planar spin network
with 10000 vertices. The graph shows the log-log distributions of
avalanche for $c_{max}=10, 30, 100, 1000, 10000$ respectively, and
the fitted line shows the power-law relation: $P(s)\propto
s^{-2.17}$. The corresponding distributions of area are in the
log-log subplot.} \label{Fig.4}
\end{figure}

As shown in Fig. \ref{Fig.2}, the distribution of the size of
avalanches is found to be
\begin{equation}
P(s)\propto s^{-\alpha}, \notag
\end{equation}
with the power-law exponent, $\alpha \approx 2.06$ (in
\cite{Ansari}, $\alpha \approx 3.3$). Moreover, the power-law
exponent, $\alpha$, is slightly affected by the finite-size effect.
For instance, $\alpha \approx 2.17$ for the spin network with
vertices 10000 in Fig. \ref{Fig.4}.

Strictly speaking, the area of avalanches is also need to be a
power-law distribution in a SOC model. The log-log subplot in Fig.
\ref{Fig.2} indicates a good power-law relation of area
(corresponding to the distribution of size of avalanche in Fig.
\ref{Fig.2}). The concurrence of the power-law relations of the size
and the area in their distributions supports that the evolution
process under the propagation rule is self-similar, which means that
the system is a SOC system.

Besides the distribution of the size and area of avalanche, the
fraction of the flat triangles (abbr. FFT) and the average color
(abbr. Avgc) of network are also of interests and studied in the
literatures \cite{Ansari,Borissov}.

A flat triangle is dual to a flat vertex which has special edge
colors making one of the three conditions in Eq. (\ref{gnc1})
saturated. The flat triangle plays a pivotal role in the evolution
of spin network. Its next evolution by the addition or subtraction
of loop flux will likely lead to a gauge-non-invariant state and
cause an avalanche of recovering gauge invariant state. Especially,
when the system is in the critical state, the fraction of the number
of the flat triangle in time should be a fixed value \cite{Ansari}.
Such result can be found in Fig. \ref{Fig.3}a and \ref{Fig.3}b. In
this model, the fraction of flat triangles rapidly drops from a
finite value ($\sim 0.165$) and saturates around a fixed value,
about $0.026$.

The average color of the network in time fits well onto a straight
line after the first relaxation period in Fig. \ref{Fig.3}(b), in
agreement with the result in \cite{Ansari} (through the repeated
work of \cite{Ansari} on the spin network with $361$, $10000$ and
$102400$ vertices respectively, we find out that the strong
oscillation of the Avgc of the network presented in \cite{Ansari} is
mainly caused by the limited size of the spin network). Since the
color is proportional to the length of the side of the triangulated
space, we are also observing the linear expansion of the universe
described by the spin network model. It is another evidence of the
statement that an asymptotic system may also display SOC
\cite{Ansari,asymptote}.

In our detailed evolving rule, the evolution of the spin network is
not independent from the initial maximum color $c_{max}$. We did
simulations for $10^7$ steps on a spin network with 10000 vertices
for $c_{max}=10, 30, 100, 1000$ and $10000$, respectively. The
distributions of size and area of avalanche for each typical run are
shown in Fig. \ref{Fig.4}. Interestingly, the distributions of size
(area) of avalanche for each $c_{max}$ very well overlap on each
other. In other words, $c_{max}$ have no obvious effect on the SOC
behavior in this model.

\section{Summary and discussion}

\label{Sec.5}

Based on the work of Borissov and Gupta \cite{Borissov}, and the
work of Ansari and Smolin \cite{Ansari}, we studied the evolution of
a frozen spin network, in which the colors of the edges (namely
their twice spins) change but the underlying graph remains fixed.
Two different classes of evolution rules are considered. One class
(the random edge model) involves choosing an edge randomly and
changing its color by an even integer. The other class (the random
vertex model) involves choosing a vertex at random and changing the
colors of its three incident edges by an even integer. Since the
changes of colors generally lead to the violation of the gauge
invariance conditions, a detailed propagation rule is defined for
the spin network to regain the gauge invariance during the
evolution. The possible propagation rules can be defined in
different ways. In the previous works \cite{Ansari, Borissov}, the
value of the color change during the evolution of the spin network
is fixed. In the present work, considering the fact that the
vertices are interdependent, we suggested that the change of color
is a variable dependent on the colors of the vertex itself and the
adjacent vertices. To be more specific, the change of color is an
even integer chosen between $2$ and $c_{max}$ (the maximum color
appearing in the initial state), with a probability distribution
dependent on the colors of the considered vertex and its neighboring
vertices. We applied this rule to the two models mentioned above
respectively, in the framework of a two-dimensional planar open spin
network. The random edge model under our propagation rule exhibits
the SOC behavior. But we had not found evidence of SOC in the random
vertex model. Furthermore, the increase of the average color of the
spin network in time exhibits the inflation of universe.

In summary, we studied the loop quantum gravity using a statistical
physics method. The results show that a quantum gravity system
(under the limit of low energy) is equivalent to a non-equilibrium
statistical system whose temporal evolution exhibits the
self-organized criticality. Moreover, our research shows that the
size of a gravity system (which also can be regarded as a universe)
expands in time. Even though the accelerated expansion of the
universe can not be proved yet in the paper, we expect that it can
be observed under more realistic and accurate propagation rules. Our
work is only the inchoate investigations of the SOC in quantum
gravity. Further works along this line, such as the dynamics of spin
network, generalization to higher dimensional space and different
propagation rules, will hopefully help us gain more insight into the
self-organization criticality in quantum gravity.

\acknowledgments

The work was supported by the National Natural Science Foundation of
China (No. 10375008, 10503001), and the National Basic Research
Program of China (2003CB716302).

\end{document}